\documentclass[draft]{agujournal2019}
\usepackage{url} 
\usepackage{lineno}
\usepackage[finalnew]{trackchanges} 
\addeditor{R1}
\addeditor{R2}
\addeditor{R3}
\addeditor{authors} 
\usepackage{soul}
\usepackage{amsmath}
\draftfalse

\journalname{JAMES}

\begin{document}

%
%


\title{\change[R2]{Using}{Finetuning} AI Foundation Models to Develop \change[R2]{Climate Model}{Subgrid-Scale} Parameterizations: \add[R2]{A Case Study on Atmospheric Gravity Waves}}

%
%




\authors{Aman Gupta\affil{1}, Aditi Sheshadri\affil{1}, Sujit Roy\affil{2,3}, Johannes Schmude\affil{4}, Vishal Gaur\affil{3}, Wei Ji Leong\affil{5}, Manil Maskey\affil{2}, Rahul Ramachandran\affil{2}}


\affiliation{1}{Department of Earth System Science, Stanford University, Stanford, USA}
\affiliation{2}{Earth System Science Center, The University of Alabama in Huntsville, Huntsville, AL, USA}
\affiliation{3}{NASA Marshall Space Flight Center, Huntsville, AL, USA}
\affiliation{4}{IBM Research, Yorktown, NY, USA}
\affiliation{5}{Development Seed, Washington, DC, USA}





\correspondingauthor{Aman Gupta}{ag4680@stanford.edu}



\begin{keypoints}
\item AI weather foundation models can be fine-tuned to create machine learning (ML) climate model parameterizations for atmospheric gravity waves 
\item The fine-tuned parameterization beats existing ML benchmarks, predicting \remove[R3]{physically }accurate wave fluxes and variability 
\item Despite predicting accurate monthly averages and strong wave events, ML models continue to struggle with the prediction of small flux values 
\end{keypoints}

%
%

%
%

\begin{abstract}
Global climate models parameterize a range of atmospheric-oceanic processes like gravity waves, clouds, moist convection, and turbulence, \add[R1]{that cannot be sufficiently resolved}. These subgrid-scale \change[R1]{parameterizations}{closures for unresolved processes} are a leading source of model uncertainty. Here, we present a \remove[R1]{radically} new approach to developing machine learning parameterizations of small-scale \change[R1]{earth}{climate} processes\remove[R1]{, i.e.,} by fine-tuning a pre-trained AI foundation model (FM). FMs are largely unexplored in climate research. A pre-trained encoder-decoder from a 2.3 billion parameter FM (NASA and IBM Research's Prithvi WxC) --- which contains a latent probabilistic representation of atmospheric evolution --- is fine-tuned (or reused) to create a deep learning parameterization for atmospheric gravity waves (GWs). \change[R1]{The parameterization learns GW effects from high-resolution, ``GW-resolving" climate data to be able to represent them in coarse-resolution climate models}{The parameterization captures GW effects for a coarse-resolution climate model by learning the fluxes from an atmospheric reanalysis with 10 times finer resolution}. A comparison of monthly averages and instantaneous evolution with a \add[R1]{machine learning model} baseline (an Attention U-Net) reveals superior predictive performance of the FM parameterization throughout the atmosphere, even in regions excluded from pre-training. This performance boost is quantified using the Hellinger distance, which is 0.11 for the baseline and 0.06\remove[R1]{ (roughly half)} for the fine-tuned model. Our findings emphasize the versatility and reusability of FMs, which could be used to accomplish a range of atmosphere- and climate-related applications, leading the way for the creation of observations-driven and physically accurate parameterizations for more earth-system processes.
\end{abstract}

\section*{Plain Language Summary}
Climate models struggle to accurately capture the physical effects of small-scale atmospheric processes like gravity waves, turbulence, and clouds, which are critical to accurately predicting future climate. These processes evolve on scales finer than typical model grid resolutions. As a result, they continue to rely on approximations, known as physical parameterizations, to represent their missing effects. The use of parameterizations introduces uncertainty and makes climate predictions less reliable. Here, we propose a new approach to improving these parameterizations using modern advances in deep learning. Specifically, we use Prithvi WxC, a large AI model trained on multiple decades of one reanalysis, and fine-tune it using limited years of gravity wave data from another reanalysis to develop an emulator capable of predicting a physically consistent atmospheric gravity wave flux evolution. The novel approach of leveraging a large AI model pre-trained on vast volumes of atmospheric data and augmenting it with limited process-specific data allows the creation of compact and easily trainable data-driven physical parameterizations. While we focus on gravity waves, our approach is flexible and can be generalized to developing data-driven parameterizations of other earth system processes.


%
%

%


%
%
%
%

\section{Introduction}
Accurate prediction of future climate is a trillion-dollar challenge with critical consequences for the world economy, food security, global health, and urban planning. Currently, state-of-the-art climate projections are highly uncertain, and much of the inherent model uncertainty stems from approximations made in subgrid-scale parameterizations \cite{Morrison.Lawrence2020, Lee.etal2023}. For instance, it has been suggested that model uncertainty accounts for 98\% of the total uncertainty in precipitation projections \cite{Wu.etal2022}. This study aims to demonstrate the untapped potential of AI foundation models (FMs) to improve traditional numerical climate models by targeting subgrid scale parameterizations.

\begin{figure}[!ht]
\centering
\includegraphics[width=0.9\textwidth]{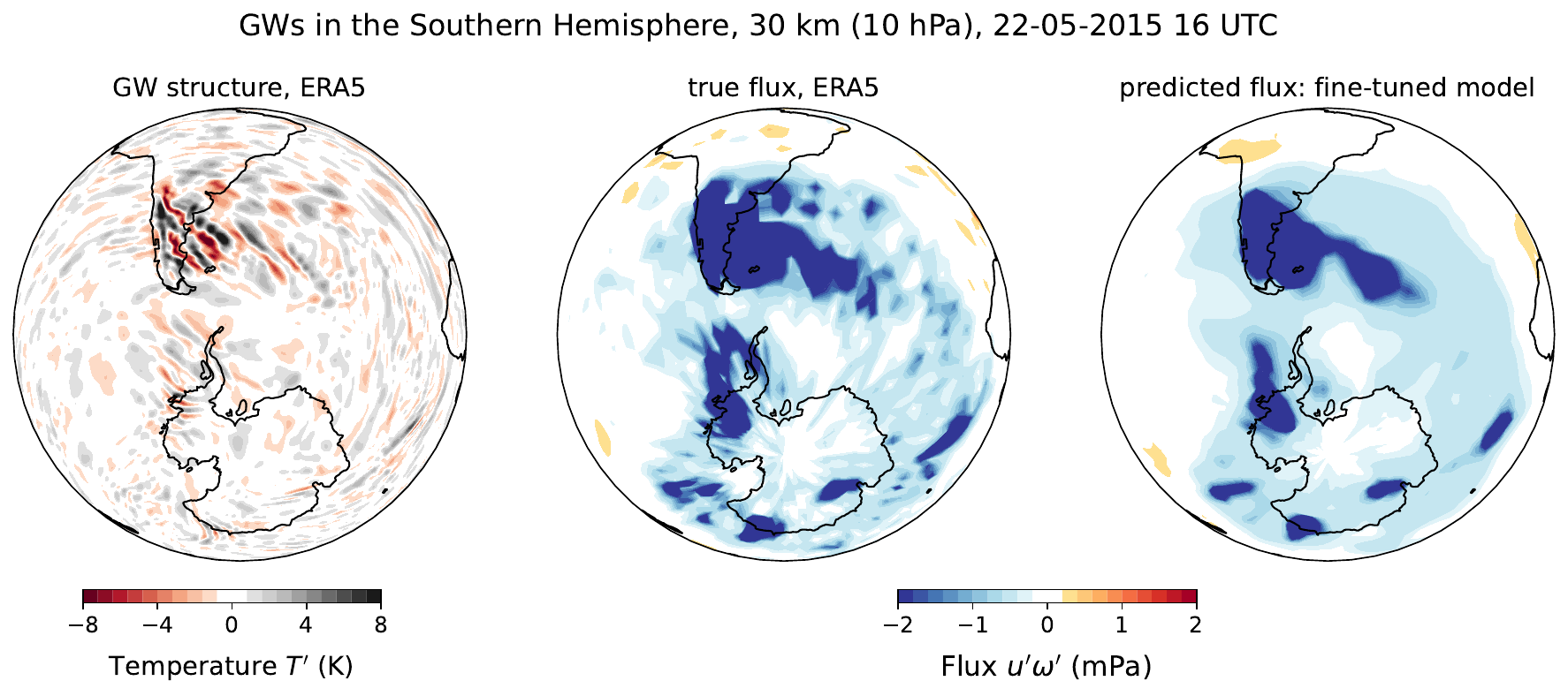}
\caption{Predictions from the fine-tuned gravity wave parameterization. The left plot shows the temperature structure of GWs over the \change[R3]{Andes, Antarctica and the Southern Ocean}{Drake Passage}, as seen in ERA5 reanalysis \cite{Hersbach.etal2020}. \add[R2]{Temperature perturbations $T'$ were computed by removing the large scales, here defined as the first 21 total wavenumbers}. The middle and right plots show the true and predicted momentum flux carried by the waves. \add[R2]{30 km is an approximate representative height since the fluxes are evaluated on a pure pressure level}. \add[R3]{Here (and throughout the study), ``true" flux refers to the flux derived from the ERA5 reanalysis, and the predicted flux is the prediction from the machine learning models trained on ERA5. Almost all gravity waves in ERA5 are model-generated. Therefore, the gravity wave structure and the inferred flux might not be a precise representation of the actual atmospheric conditions.}}
\label{fig:intro}
\end{figure}

FMs can be broadly defined as task-agnostic large AI models which are pre-trained using a self-supervised learning objective \cite{Bommasani.etal2022}, such as weather forecasting from time $t$ to $t + \Delta t$. Given the single-task limitation of existing AI weather forecasting models \cite{Lam.etal2023, Bi.etal2023, Price.etal2025}, despite their massive compute requirements, FMs are developed to be versatile and present the next frontier in AI research. Pre-trained FMs are subsequently fine-tuned to perform a broad range of sub-tasks, a.k.a., downstream tasks. FMs are largely unexplored in climate science, and only a couple of weather and geospatial FMs exist to date: AtmoRep \cite{Lessig.etal2023}, Aurora \cite{Bodnar.etal2025}, and Prithvi \cite{Jakubik.etal2023}. To our knowledge, only weather-related downstream applications of FMs have been explored thus far, including hurricane track and intensity prediction, air quality predictions, downscaling, vegetation burn-scar detection, etc. \remove[R1]{The limited stability of auto-regressive FM rollouts over longer timescales has sown doubt over their utility for climate applications.}

Here, we use a recently developed, state-of-the-art FM, Prithvi WxC \cite{Schmude.etal2024} (hereafter Prithvi), to demonstrate that FMs can be used to develop deep-learning parameterizations for unresolved earth system processes for climate modeling applications. The parameterization for atmospheric gravity waves presented here is capable of representing the missing effects of atmospheric GWs in global climate models. We blend the pre-trained encoder-decoder pair from Prithvi with high-resolution gravity wave momentum flux data (see Methods) to create a fine-tuned AI model that skillfully predicts subgrid-scale GW activity and outperforms \change[R3]{leading}{existing} benchmarks \cite{Gupta.etal2024} for deep-learning-based GW flux prediction. The study motivates and calls for the strategic use of FMs for climate-related tasks by showing how to leverage observations and FMs to efficiently achieve predictive tasks that might otherwise require much larger volumes of training data.

Atmospheric GWs are ubiquitous multiscale (spatial scale $\mathcal{O}$(1)-$\mathcal{O}$(1000) km) oscillations generated \change[R1]{around thunderstorms, jet stream disturbances, flow over mountains, etc.}{by atmospheric convection, jet stream disturbances, geostrophic imbalance, and flow over mountains} \cite{Fritts.Alexander2003}. GWs dynamically couple\remove[R1]{the} different layers of the atmosphere by carrying near-surface momentum and energy to stratospheric and mesospheric heights. \add[R1]{In the troposphere, gravity waves play a critical role in setting the location and strength of the jet streams} \cite{Palmer.etal1986}.
In the stratosphere, they influence the quasi-biennial oscillation (QBO) of tropical winds \cite{Giorgetta.etal2002}, the springtime breakdown of the Antarctic polar vortex \cite{Gupta.etal2021a}, and thus Antarctic surface temperatures \cite{Choi.etal2024}. \add[R3]{In the mesosphere, GWs are the primary driver of the pole-to-pole overturning circulation.} \cite{Becker2012}. GW-induced cold anomalies in the polar winter stratosphere provide suitable conditions for the formation of polar stratospheric clouds, enabling reactions that promote the destruction of ozone \cite{Doernbrack.etal1999, Hoepfner.etal2006, Hoffmann.etal2017}. 
Aside from their influence on climate variability, GW-induced turbulence can influence commercial air travel and is believed to have caused the sudden plunging of Singapore Airlines flight SQ321 on 21 May 2024 \cite{Hirschfeld2024}.

The current climate model grid resolution \change[R1]{(100-200 km)}{(50-100 km)} is insufficient to fully resolve dynamically important processes like gravity waves, clouds, and turbulence. The traditional approach to represent these missing processes has been to couple the numerical fluid solver with a suite of \emph{sub-grid scale parameterizations} to approximately capture the unresolved effects of these processes \cite[to name a few]{Alexander.Dunkerton1999, Lott.Miller1997, Bogenschutz.etal2012, Iacono.etal2000}.

Parameterizations are often not well constrained by observations and, for computational reasons, contain a series of simplifications that compromise their physical accuracy.  For GWs, these often include an idealized source spectrum and, generally, complete neglect of their transient evolution and horizontal propagation  \cite{Achatz.etal2024}. Further, their parametric tuning is often sub-optimal because the parameters are optimized to replicate only certain atmospheric features of interest. These inductive biases \add[R1]{(due to simplifying assumptions)} often add up and result in inaccurate model dynamics, such as the prominent ``cold-pole bias" \cite{McLandress.etal2012a}, leading to large uncertainties in future climate projections \cite{Golaz.etal2013, Mauritsen.etal2012, Zhao.etal2018}.

Data-driven approaches are increasingly being used to develop fast \add[R1]{GWF}{gravity wave flux (GWF)} emulators for climate models of varying complexity \cite{Chantry.etal2021,Espinosa.etal2022,Sun.etal2024,Hardiman.etal2023,Connelly.Gerber2024,Lu.etal2024, Ukkonen.Chantry2024}. \add[authors]{These emulators complement existing efforts to develop nonlocal GW parameterization using physics-based approaches} \cite{Voelker.etal2023, Eichinger.etal2023}. Despite being effective, since these emulators are trained on parameterization data itself, they do not offer an improved process physics representation. Here, we fine-tune the FM on \emph{resolved} GWFs. Training on resolved GWFs allows the \change[R2]{NNs}{neural networks (NNs)} to learn key physical \change[R2]{properties}{effects} of GWs directly from fine-tuning data datasets.

Our fine-tuned parameterization, created by blending Prithvi and \add[R1]{(partial GW resolving)} ERA5 reanalysis skillfully predicts the GW momentum fluxes for a provided background atmospheric state (as shown in Fig \ref{fig:intro}). The GW structure on \change[R2]{24}{22} May 2015 and the momentum flux carried by the waves is shown in Fig \ref{fig:intro}. The fine-tuned model \change[R1]{correctly predicts the intermittent intensification of}{accurately predicts} the fluxes over the \change[R3]{Andes in South America, the Amundsen Sea and Ross Sea in Antarctica,}{Drake Passage and the Southern Ocean.} The fluxes over the Andes extend\add[R1]{ sufficiently} leeward (up to 80 degrees longitude) to over the Southern Ocean, indicating that the fine-tuned model can learn and represent the lateral propagation and transient evolution of the generated waves; a physical feature absent in most current GW parameterizations \cite{Plougonven.etal2020}.

This fine-tuned parameterization for GWs can be coupled to a coarse-resolution climate model to represent ``missing" GW physics. Since Prithvi was pre-trained on key atmosphere-ocean-land variables, the scope of this approach transcends GWs, and fosters and expedites the creation of physically accurate AI parameterizations of other small-scale earth system processes, ultimately contributing to the development of accurate and interpretable hybrid climate prediction systems.

\section{Methods}

\subsection{The Prithvi WxC Foundation Model for Weather and Climate}
Prithvi WxC, jointly developed by NASA and IBM, is a transformer-based deep learning architecture that combines ideas from several
recent transformer architectures in order to effectively process regional and global dependencies of
the input data and to efficiently process longer sequence lengths of tokens. \add[R1]{Any image input to a transformer is broken down into smaller square patches that are then projected to a higher-dimensional space to represent the image in numerical space. These projections, which represent discrete amounts of information, are referred to as a token}. This allows the model to,
for instance, run in different spatial contexts or infuse additional tokens \add[R3]{(adding more information as tokens into later stages of the model instead of the input to preserve or enhance context)} from off-grid measurements
into the model during fine-tuning. Prithvi has 2.3 billion trainable parameters and is trained on 160 atmospheric channels using 40 years of 3-hourly MERRA-2 reanalysis \cite{Gelaro.etal2017} data at a 0.5$^{\circ}$$\times$0.625$^{\circ}$ spatial resolution. \add[R3]{The channels include 20 surface variables (winds, pressure, latent heat flux, surface roughness, etc.) and 10 atmospheric variables (winds, clouds, humidity, etc.) on 14 vertical pressure levels each. These variables are tabulated in Table 2 and Table 3 in Appendix A of} \cite{Schmude.etal2024}.

The validation of Prithvi extends from zero-shot evaluations for reconstruction and forecasting
to other downstream tasks, such as the downscaling of weather and climate models, the prediction of hurricane
tracks, and climate model parameterization. The architecture of the pre-training backbone is shown in Fig \ref{fig:merged_attnunet_fmencdec}a. \add[R1]{As shown in the figure, Prithvi was trained on a masked reconstruction objective, which means that in addition to minimizing the root mean square for the predictions, the model also minimized reconstruction error from masked input data. A fixed fraction (50\%) of the input cells were masked, and the model was tested on how well it could fully reconstruct the global field from the masked data}. More details are provided in \citeA{Schmude.etal2024}\add[R1]{, where Equation 1 and Section 2.5.1 focus on masked reconstruction}.

\add[R3]{The fine-tuning task presented in this manuscript is identical to that presented in Section 3.2 of} \citeA{Schmude.etal2024}\add[R2]{, i.e., the ``Climate Model Parameterization for Gravity Wave Flux" task.} \cite{Schmude.etal2024} \add[R3]{only briefly showcase it as one among many applications of an AI foundation model, but here, we delve deeper and provide a more detailed analysis of the task.}

\subsection{Preparing Training Data for GW Flux Prediction}


The fine-tuning data for GW flux prediction was prepared using ERA5 global reanalysis data \cite{Hersbach.etal2020} retrieved at 25 km horizontal resolution, 137 vertical levels, and an hourly frequency. At this resolution, ERA5 resolves GWs with wavelengths longer than 150-200 km, providing global, multi-decadal information on atmospheric GW evolution. \remove[R1]{Practically none of these waves are resolved by a typical climate model.} \add[R2]{We note that the most accurate way to report the accuracy of spectral-based climate models, however, is to report the spectral resolution and the native grid resolution. ERA5 uses 639 spectral wavemodes on a sphere to represent the variables and iteratively transforms variables between the spectral space and a N320 reduced Gaussian grid with $\sim$31 km resolution (which has a different number of longitudes for different latitudes to maintain equal spacing of points in both the tropics and the polar regions). Since the model outputs are interpolated and publicly presented on a regular 25 km latitude-longitude grid, we hereafter refer to ERA5 as having a 25 km resolution.}

ERA5 does not provide GW momentum fluxes. So, we compute the fluxes by applying Helmholtz decomposition (HD) \cite{Lindborg2015, Kohler.etal2023} on the raw ERA5 output as follows. First, the horizontal winds ($u$ and $v$) from ERA5 are decomposed into rotational and divergent components:
\begin{equation}
    \vec{u} = (u,v) = -\nabla\phi + \nabla\times\psi
\end{equation}
where $\phi$ is the potential function such that $\nabla\phi$ is irrotational. Similarly, $\psi$ is the rotational streamfunction function such that $\nabla \times \psi$ is non-divergent. $\phi$ and $\psi$ are used to reconstruct the divergent (div) and rotational (rot) parts of the horizontal flow as:
\begin{equation}
    \vec{u} = (u,v) \overset{HD}{\longrightarrow} (u_{div},v_{div}) + (u_{rot}, v_{rot}).
\end{equation}
\add[R2]{Next, to ensure that the large-scale background (including divergent equatorial Kelvin wave components) is completely removed from the divergent flow, an additional fixed-wavenumber high-pass filter is applied by removing the T21 truncated divergent velocity, $(u_{div,T21},v_{div,T21})$, from the divergent flow. This operation is expressed as:}
\begin{equation}
    (u'_{div},v'_{div}) = (u_{div} - u_{div,T21}, v_{div} - v_{div,T21})
\end{equation}
These are \change[authors]{combined}{multiplied} with the zonal mean removed \change[R3]{vertical}{pressure} velocity \add[R3]{anomaly} ($\omega'$) to compute the directional GW momentum fluxes: 
\begin{equation}
    \vec{F} = (F_x,F_y)  = g^{-1}(u'_{div}\omega',v'_{div}\omega').
    \label{eqn:hd_fluxes}
\end{equation}
which we aim to learn using the \change[R2]{ML}{machine learning (ML)} models. Here, $g$ = --9.81 m/s$^2$ is the acceleration due to gravity. \add[R2]{Hereafter, we use the shorthand notation $u'\omega'$ and $v'\omega'$ to denote the directional fluxes in Equation }\ref{eqn:hd_fluxes}.

The procedure is applied to create the fine-tuning training data. The top 15 of the 137 vertical levels are discarded due to artificial model damping. All input-output pairs are coarse-grained from a 25 km resolution to a 64 latitudes $\times$ 128 longitudes grid (roughly 2.8$^{\circ} \approx$ 280 km resolution in the tropics) to obtain conservative wave averages. The fluxes are computed for four years: 2010, 2012, 2014, and 2015. This corresponds to roughly 35k training\add[R2]{+validation} samples \add[R2]{since one 64 $\times$ 128 $\times$ 122 hyperslab makes up 1 training sample}, which could be considered ``data-scarce".

\begin{figure}[!h]
\centering
\includegraphics[width=0.85\textwidth]{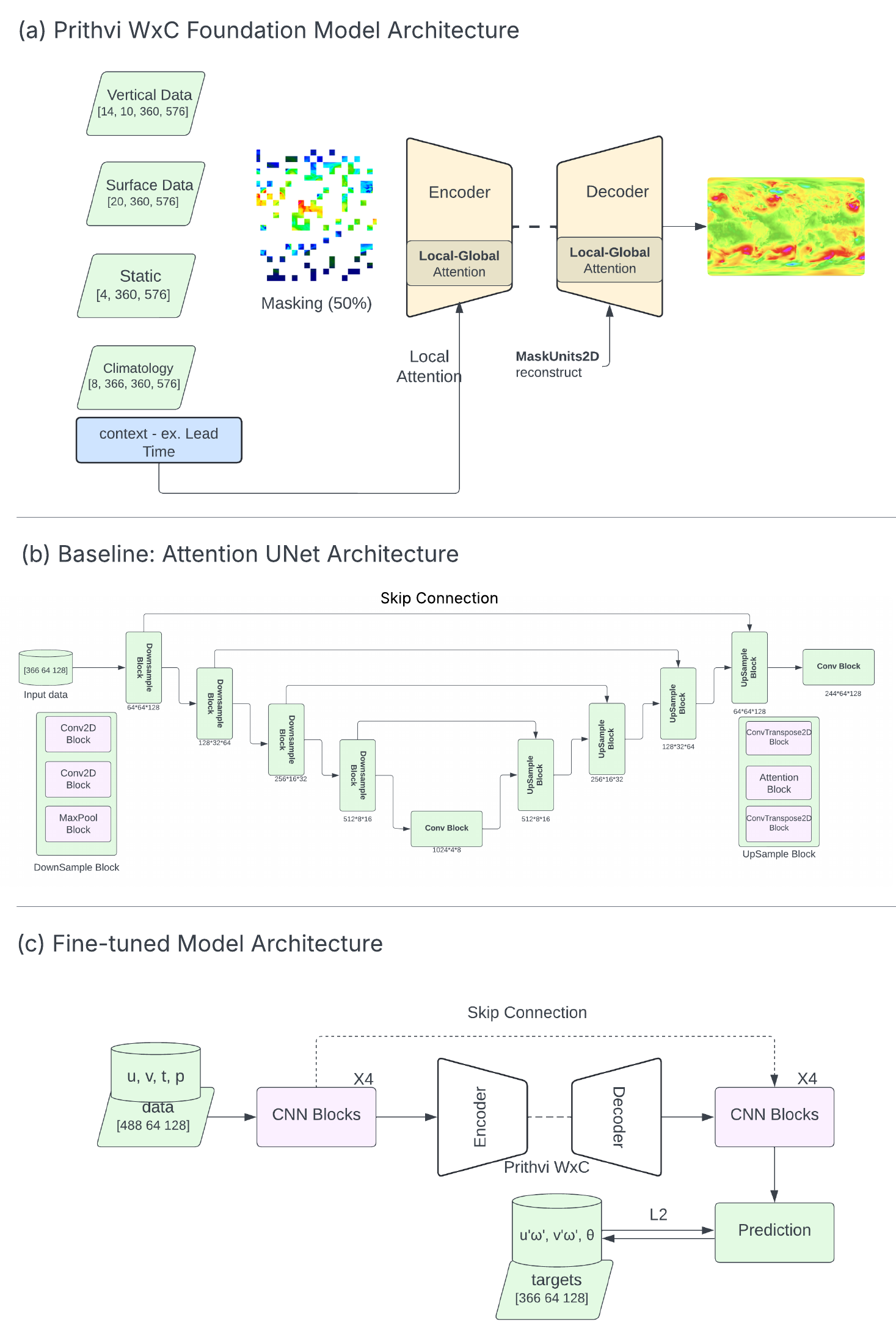} 
\caption{(a) Pre-training model architecture for Prithvi WxC. The encoder and decoder blocks from Prithvi WxC are frozen and used for fine-tuning. \add[R3]{10 atmospheric variables on 14 vertical levels, 20 surface variables, 4 static variables, and 8 climatology variables for 366 days in a year, all on a 360 (lat) x 576 (lon) grid, for the input.}(b) Model Architecture for attn unet \add[R3]{using 3 input variables, each on 122 vertical levels and a 64 (lat) x 128 (lon) grid} (schematically identical to \citeA{Oktay.etal2018}). (c) The FM fine-tuning architecture comprises (in order) 4 learnable convolutional layers, the frozen encoder, the frozen decoder, and 4 more learnable convolutional layers. A skip connection connects the former and latter convolutional layers. \add[R3]{Takes four input variables, each on 122 vertical levels, and on a 64 (lat) x 128 (lon) grid.} \add[R3]{The blue block in the bottom left in (a) refers to the additional infused context or relevant information added at later stages in the hidden layers, for example, the lead time at which the FM makes the predictions. B refers to the minibatch size, that is, the number of samples randomly sampled from the full input to train the model for a single iteration/forward pass.}}
\label{fig:merged_attnunet_fmencdec}
\end{figure}

\textbf{Variables for \change[R1]{baseline}{training the U-Net}:} the input consists of winds $u$, $v$, potential temperature $\theta$, which is a function of temperature $T$ and pressure $p$ (in hPa) as $\theta=T(p/1000)^{-0.286}$, each on 122 vertical levels, 64 latitudes and 128 longitudes. Similarly, the output comprises fluxes $u'\omega'$ and $v'\omega'$, each on 122 vertical levels, 64 latitudes, and 128 longitudes (Fig \ref{fig:merged_attnunet_fmencdec}b).

\textbf{Variables for fine-tuning \add[R1]{the FM}:} this is slightly different from the baseline. The fine-tuning input consists of winds $u$, $v$, temperature $T$, and pressure $p$ \add[R1]{(instead of $u$, $v$, and $\theta$)}, each on 122 vertical levels, 64 latitudes, and 128 longitudes. Similarly, the outputs are potential temperature $\theta$ (for validation) and fluxes $u'\omega'$ and $v'\omega'$ \add[R1]{(instead of just $u'\omega'$ and $v'\omega'$)}, each on 122 vertical levels, 64 latitudes and 128 longitudes (Fig \ref{fig:merged_attnunet_fmencdec}c).

\add[R1]{\textbf{Variable Normalization:} Each variable is normalized differently. The zonal wind $u$ is normalized as: $u \rightarrow (u-u_{mean})/u_{std}$, where $u_{mean}$ and $u_{std}$ are the global mean and standard deviation. Similarly for $v$ and $t$. Pressure was scaled as $p \rightarrow log_{10}(p)$, and potential temperature was scaled as $\theta \rightarrow \theta/1000$. Lastly, global mean and global standard deviations of $u'\omega'$ were used to scale the flux as $u'\omega' \rightarrow \left[(u'\omega' -u'\omega'_{mean})/u'\omega'_{std} \right]^{1/3}$}

\add[R2]{\textbf{A note on limited GW representation in ERA5:} All models considered in this study are trained on resolved wave fluxes from ERA5, and so the objective of the models is to reproduce the ERA5 fluxes as accurately as possible. For this reason, the fluxes from ERA5 are occasionally referred to as ``true" fluxes, since they comprise the training and validation set. 

ERA5 provides multi-decadal atmospheric coverage at a moderately high resolution; however, we caution against the limited GW representation in ERA5, due to which GW fluxes in ERA5 might not be a true representation of the actual GW fluxes in the atmosphere. Multiple recent studies have validated both substantial similarities and systematic differences between GWs in ERA5 and GWs in high-resolution models and observations} \cite{Pahlavan.etal2023, Yoshida.etal2024, Lear.etal2024, Gupta.etal2024d}\add[R2]{, which could be rooted in multiple factors. Firstly, with a resolution of 25 km, ERA5 does not resolve a portion of the atmospheric GWs with wavelengths shorter than 150 km. These waves could make notable contributions to the large-scale atmospheric circulation} \cite{Polichtchouk.etal2022, Polichtchouk.etal2023}\add[R2]{. Secondly, while the large-scale winds and temperature might be constrained to some degree, all GWs in ERA5 are model-generated in response to the constrained background state. Thirdly, known biases in precipitation,  clouds, land, and upper surface winds can result in biased GW generation in response to changes in these fields. This can be particularly important for small-scale convectively generated GWs which have a wide phase spectrum, and are likely to transport the momentum to mesospheric heights before dissipation} \cite{Kim.Chun2015, Achatz.etal2024}. \add[R2]{Lastly, the use of a hydrostatic dynamical core to produce ERA5 means a compromised representation of non-hydrostatic GWs, potentially leading to an incorrect wave aspect ratio for a given angular frequency.}

\add[R2]{Such differences could also exist among identically initialized high-resolution models with different underlying numerics, as noted by} \citeA{Stephan.etal2019}, \citeA{Kruse.etal2022}, and \citeA{Prochazkova.etal2023}. 

\add[R2]{We have coarsegrained all the fluxes from the 25 km grid to a coarser 280 km grid, which is more consistent with a coarse-climate model's operational resolution. This serves two purposes. 
First, this reduces the GW flux estimation problem to a simpler problem - that of learning wave-averaged fluxes, and not learning phase-dependent fluxes associated with the different phases of the waves (which, arguably, would be more challenging). Second, since a climate model with a 280 km grid would generate practically no GWs, coupling this ML scheme to a coarse climate model would ensure that all the GW tendencies are supplied by the ML scheme itself.}

\subsection{Baseline Model}

An advanced baseline was created by training an Attention U-Net model (hereafter attn unet) \cite{Oktay.etal2018} on the ERA5 data. The input is downsampled using four convolutional blocks and then upsampled using four convolutional blocks. The skip connection at each level comprises learnable attention layers. For every downsample (upsample), the number of channels increases (decreases) by a factor of 2 but all spatial dimensions reduce (increase) by a factor of 2. As a result, the baseline model consists of over 35 million learnable parameters and provides a robust comparison benchmark for the fine-tuning model. The learning rate for the model was set to 10$^{-4}$. On a single A100 80 GB GPU, the model took around 110 hours to complete 100 epochs.


\subsection{Designing the Fine-tuning Model}

The architecture schematic for the fine-tuning is shown in Fig \ref{fig:merged_attnunet_fmencdec}c. During fine-tuning, we freeze the encoder and decoder from Prithvi WxC. The frozen encoder is preceded by 4 learnable convolutional blocks each with an increasing number of hidden channels, i.e., $C$, 2$C$, 4$C$, and then 8$C$, where $C$ = 160. Likewise, the frozen decoder is succeeded by 4 new learnable convolutional blocks. \change[R3]{Since gravity wave flux prediction is an instantaneous flux calculation task}{For instantaneous prediction of gravity wave fluxes}, we fix Prithvi's lead time $\delta t$ to zero. The instantaneous model input for fine-tuning has the shape [1, 488, 64, 128] where the 488 channels comprise the four background variables $u$, $v$, $t$ and $p$ on 122 vertical levels each, and on a 64 $\times$ 128 horizontal grid, as discussed above. The model was fine-tuned to produce an output with shape [1, 366, 64, 128] comprising of the potential temperature $\theta$,  and fluxes $u'\omega'$, and $v'\omega'$ on 122 vertical levels each. The model was trained for 26 hours on 2 nodes of 4 A100 80GB each for 100 epochs\remove[R3]{, where each node had 4 A100 GPUs}. However, the model error converged to lower than the baseline model error after just 40 epochs of training.

\subsection{Training both models}

Both the models use global information as input to predict global fluxes as output. This provides a strong contrast to traditional ``single-column" climate model parameterizations. Access to the global atmospheric state allows the models to learn spatio-temporal correlations and \add[R3]{the effects of} horizontal propagation of gravity waves.

\add[R1]{Both models were trained and validated for 100 epochs on 4 years, i.e., 48 months, of ERA5 background conditions and fluxes. The baseline model's global RMSE loss dropped from an epoch 1 loss of 0.38 to plateauing near 0.17 over 100 epochs (a 40\% reduction). In contrast, the fine-tuned model showed much faster convergence, dropping from an epoch 1 loss of 0.275 to 0.16 (40\% reduction) over just 5 epochs and converging to 0.106. 

Since the main focus of the study is to highlight the application of FMs to make quick emulators, at present, only the month of May 2015 was used for validation; the remaining 47 months were used for training.
Both models leveraged a \emph{U-Net-like} architecture with skip connections to promote the extraction of high-frequency information from the source data. Both models were trained with an identical minibatch size of 4, i.e., four randomly selected timeframes of each variable formed input during a single forward and backward pass of the model. We re-emphasize that Prithvi WxC was pre-trained on the MERRA-2 dataset, but the fine-tuning was accomplished using ERA5 data instead. Both models yielded similar inference times on a single A100 GPU for an identical minibatch size of 4.}  

Both models were optimized using MSE Loss, which is defined as:
\begin{equation}
    \mathcal{L}(\vec{x},\vec{y}) = \frac{1}{n}\sum_{i=1}^n(x_i-y_i)^2
\end{equation}
where $x_i$ is the $i$-th prediction compared against the $i$-th measured sample $y_i$.

\subsection{Hellinger Distance} 
Given two probability densities, $p$ and $q$, their Hellinger distance, $\mathcal{H}$ \cite{Hellinger1909}, is defined as: 
\begin{equation}
    \mathcal{H}(p,q) = 1- \int_{x \in X} \sqrt{p(x)q(x)} dx.  
    \label{eqn:hellinger}
\end{equation}
By definition, $\mathcal{H} \in \left[0,1\right].$ A Hellinger distance of 0 means the distributions are identical almost everywhere, while a Hellinger distance of 1 implies the distributions are disjoint, i.e., $p$ is non-zero wherever $q$ is zero, and vice versa.

\add[R3]{Hellinger distance measures the statistical distance between two distributions. In Section }\ref{sec:results}\add[R3]{, Hellinger distance is used to quantify the difference (or statistical distance) between the flux distributions from ERA5 and the prediction flux distributions to estimate the quality of predictions by both the attn uNet and the fine-tuned model.}

\section{Results}
\label{sec:results}
\subsection{Instantaneous, Intermittent Evolution of Gravity Waves}

We focus on predicting $u'\omega'$, which is the vertical flux of zonal momentum carried by GWs. Its vertical derivative equals the net forcing tendency (acceleration) exerted by GWs on the zonal wind\remove[R1]{ upon dissipation}. The findings are similar for the vertical flux of meridional momentum, $v'\omega'$, and equivalent plots for $v'\omega'$ are shared in the Appendix. In all instances, the predictions are compared to both the \remove[R2]{true }fluxes from ERA5 and to predictions from the \change[R3]{established leading}{existing} benchmark model, attn unet. 

The fine-tuned parameterization generates a remarkably accurate prediction of the intermittent generation and temporal coherence of GW packets, although no explicit considerations were made to embed recurrence in the underlying fine-tuning architecture. 

\begin{figure}[!ht]
\centering
\includegraphics[width=1\textwidth]{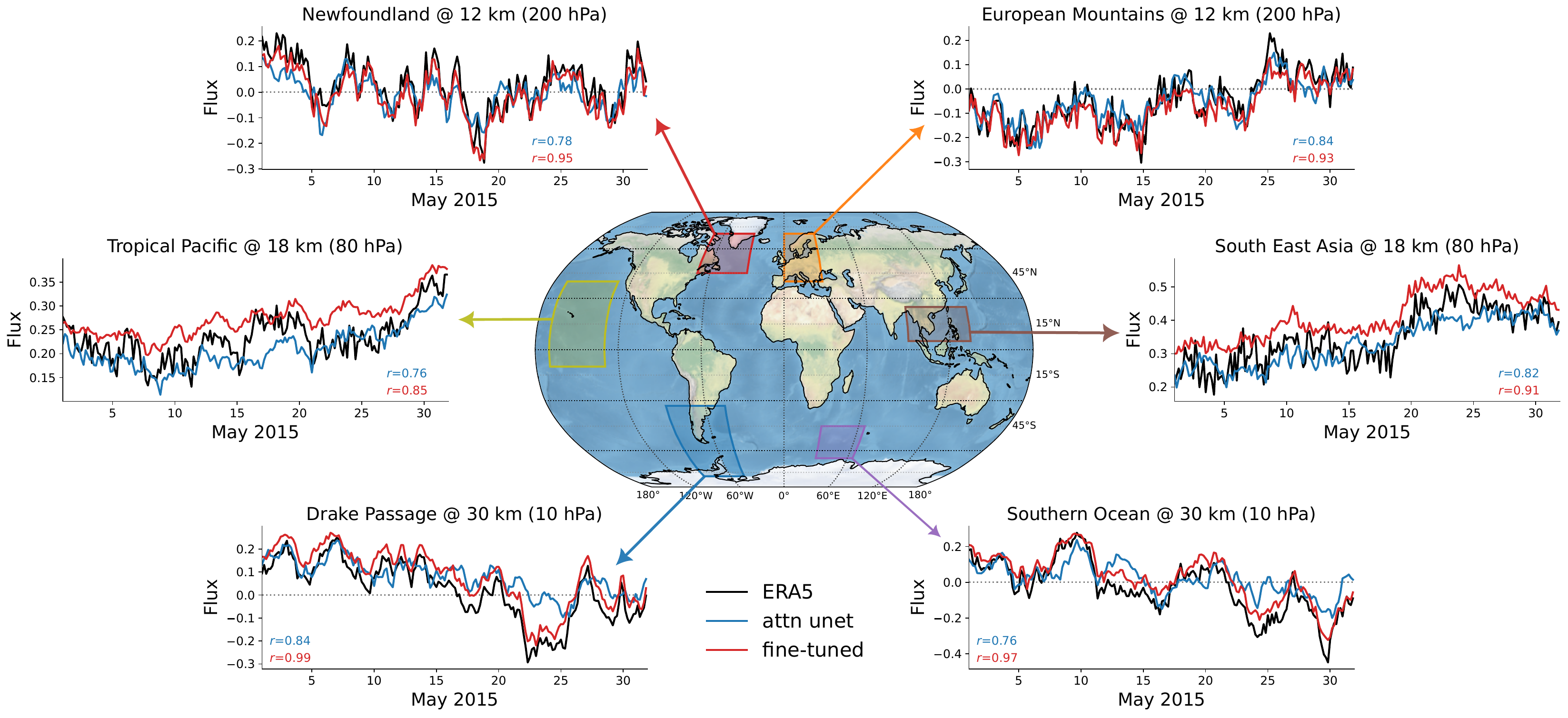} 
\caption{\remove[R2]{True i}{I}nstantaneous \add[R2]{(non-dimensional/normalized)} fluxes for May 2015 resolved in ERA5 reanalysis (black), predicted fluxes from attn unet (blue), and from the fine-tuned parameterization (red), over six well-known GW hotspots. The numbers show the respective Pearson correlation coefficients with respect to ERA5. The fluxes in the winter hemisphere are shown at 30 km, whereas the fluxes in the summer hemisphere are shown at 12 km, as GW activity in the summer stratosphere is substantially weaker. \add[R2]{30 km and 12 km are approximately representative heights since the fluxes are evaluated on pure-pressure and hybrid-pressure levels respectively, that do not equate to similar geopotential heights throughout the domains.} \add[R3]{The Pearson correlation coefficient (between ERA5 and attn unet) is computed as the covariance between the ERA5 fluxes and attn unet fluxes divided by the product of ERA5 and attn unet standard deviations.}}
\label{fig:timeseries}
\end{figure}

The time evolution of the box-averaged fluxes was analyzed for May 2015 over 6 well-known hotspots of GW activity and is illustrated in Fig \ref{fig:timeseries}. 
The three predominantly orographic hotspots \add[R3]{(Newfoundland, European Mountains, and Drake Passage)} and three nonorographic hotspots \add[R3]{(the tropical Pacific Ocean, Southeast Asia, and the Southern Ocean)} were selected based on established zonal flux climatology \cite{Hindley.etal2020, Wei.etal2022}. Nonlocal propagation of GWs is more prominent in the winter stratosphere due to a stronger vertical shear \cite{Sato.etal2012, Gupta.etal2024c}, so wherever possible, the transient evolution is shown in the upper winter stratosphere (10 hPa $\sim$ 30 km), i.e., the Southern Hemisphere for May. For regions in the summer/Northern hemisphere, the fluxes are instead analyzed in the upper troposphere (200 hPa $\sim$ 12 km).

The fine-tuned FM generates substantially better predictions over all six hotspots.
Most notably, for the Drake Passage (predominantly orographic waves) and the Southern Ocean (nonorographic waves), the Pearson correlation coefficients of the predictions from the fine-tuned model (vs. ERA5) are as high as 0.99 and 0.97, respectively. In comparison, the respective correlations for the attn unet are 0.84 and 0.76. The correlation with ERA5 is the weakest over the Tropical Pacific Ocean, but even then, the fine-tuned model has a higher correlation of 0.85, higher than attn unet's 0.76. \add[R3]{The results in the lower stratosphere are mixed. Even though the fine-tuned model has a higher correlation over the tropical box, flux magnitudes from attn unet match better with ERA5. Noisier fluxes due to tropical convective GWs with a broad range of phase speeds appear to be more challenging to predict than extratropics GWs. Expanding the feature set to include diabatic heating or precipitation-related information could potentially lead to performance gains in the region.}

\begin{figure}[!ht]
\centering
\includegraphics[width=\textwidth]{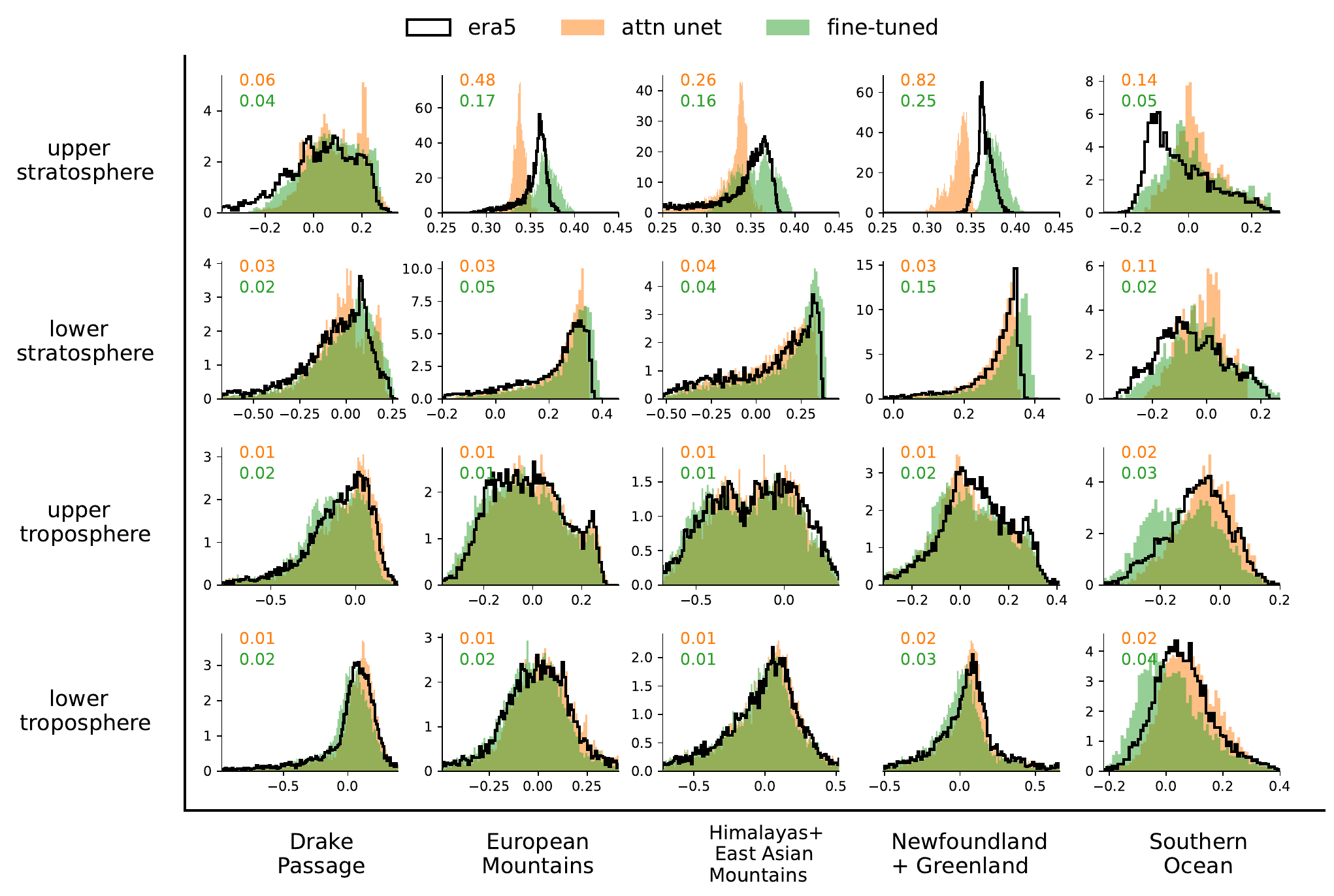} 
\caption{\add[R1]{May 2015 averaged} GW momentum flux distributions divided according to hotspots and vertical regions in the atmosphere. \add[R2]{The figure shows non-dimensional fluxes as predicted by the models for ease of comparisons. The y-axis is the distribution density}. The fluxes are averaged over the boxes outlined in Figs \ref{fig:timeseries} and \ref{fig:all_hotspots}. \add[R3]{The numbers in orange and green indicate the Hellinger distances for the time-averaged flux distributions for the attn unet and the FM model, respectively.} \add[R1]{Lower troposphere: 500 hPa to surface (0-10 km height), upper troposphere: 100-500 hPa (10-16 km height), lower stratosphere: 30-100 hPa (16-25 km height), upper stratosphere: 10-30 hPa (25-45 km height).}}
\label{fig:hotspots_histogram}
\end{figure}

The successful prediction of both the spontaneous bursts of flux intensification in the tropics (from tropical storms and convective systems) and the slower flux intensification in the midlatitudes (from mountains and storm tracks) shows that the fine-tuned model proficiently learns the intermittent excitation and horizontal evolution of medium-to-small-scale atmospheric variability directly from data. This is further corroborated by the spatial structure of the predicted flux in Fig \ref{fig:intro}, which shows that the model predicts both the fluxes over the Southern Andes and the laterally propagated fluxes in its vicinity. The wave packets preserve their coherence in time as they propagate away from their sources of excitation (see the animation provided as Supplementary Information).

\subsection{Regionwise Averaged Flux Distribution}

The dynamical evolution of atmospheric GWs can vary substantially with height (troposphere vs. stratosphere), region (latitude and longitude), season (summer vs. winter), etc. Fig \ref{fig:hotspots_histogram}, therefore, shows the predicted and \change[R2]{true}{ERA5} GW flux distributions partitioned by individual hotspots and varying atmospheric altitudes. The fine-tuned model captures the entire range of flux magnitudes over different GW hotspots (Fig \ref{fig:hotspots_histogram}). \change[R3]{In the troposphere and lower stratosphere, both the baseline and the fine-tuned model produce similar distributions.}{In the troposphere and the lower stratosphere, the models provide comparable performance. In fact, in some regions, such as the lower stratosphere over Newfoundland, and the troposphere over the Southern Ocean, the Hellinger distances are slightly better for the attn unet model.} In the upper stratosphere, however, the fine-tuned model generates a substantially more consistent distribution than the baseline. Such distributions are challenging to replicate, as the waves excited near the surface are progressively filtered and dissipated as the waves propagate to stratospheric and mesospheric altitudes. 

The Hellinger distances of the distributions for both the baseline and the fine-tuned model (w.r.t. ERA5) are shown for each hotspot and height. Ideally, a Hellinger distance of 0 indicates that the predicted distribution is identical to the \remove[R2]{true }distribution from ERA5. The fine-tuned model outperforms the baseline, yielding a lower Hellinger distance in practically all regions (the only
exception being the lower stratosphere over Newfoundland). The improvement is more evident in the upper stratosphere. Both models generate low Hellinger distances in the troposphere and most of the lower stratosphere, indicating a distribution similar to ERA5, at least in a cumulative sense. However, all regions in the upper stratosphere have higher Hellinger distances than down below, with Hellinger distances reaching up to 0.82 for the baseline over Newfoundland, revealing key biases in the summer hemisphere.

Most interestingly, the baseline model has a lower variance (and hence poorer predictive skill) than the fine-tuned model \add[R2]{in multiple stratospheric blocks}, even though Prithvi was not trained on upper stratospheric data at all (Prithvi's vertical spacing is shown in Fig \ref{fig:vert_spacing}). This highlights another benefit of using an FM's encoder-decoder that allows the creation of a consistent mapping between the FM's learned embedding space and the fine-tuning data. The performance improvement, then, can be attributed to a combination of (a) the substantially higher volumes (40+ years) of pre-training data as opposed to merely four years of ERA5 data for fine-tuning and baseline, and (b) the fine-tuning model efficiently leveraging the latent space of the pre-trained Prithvi to unify the learning from MERRA2 during pre-training and ERA5 during fine-tuning. As a result, the fine-tuning model substantially outperforms the UNet baseline even when trained on the same set of data. More simply put, despite not being trained in the upper stratosphere, training on over four decades of atmospheric data on a masked reconstruction objective (as described in Section 2.1) likely allows more consistent mappings between Prithvi's embedding space and the fine-tuning input.

A similar partition of the ERA5 and predicted monthly mean distributions, but partitioned by different latitude bands, is shown in Fig \ref{fig:regional_histogram}.

\subsection{Vertical Mean Profile and Variability}

\begin{figure}[!ht]
\centering
\includegraphics[scale=0.37]{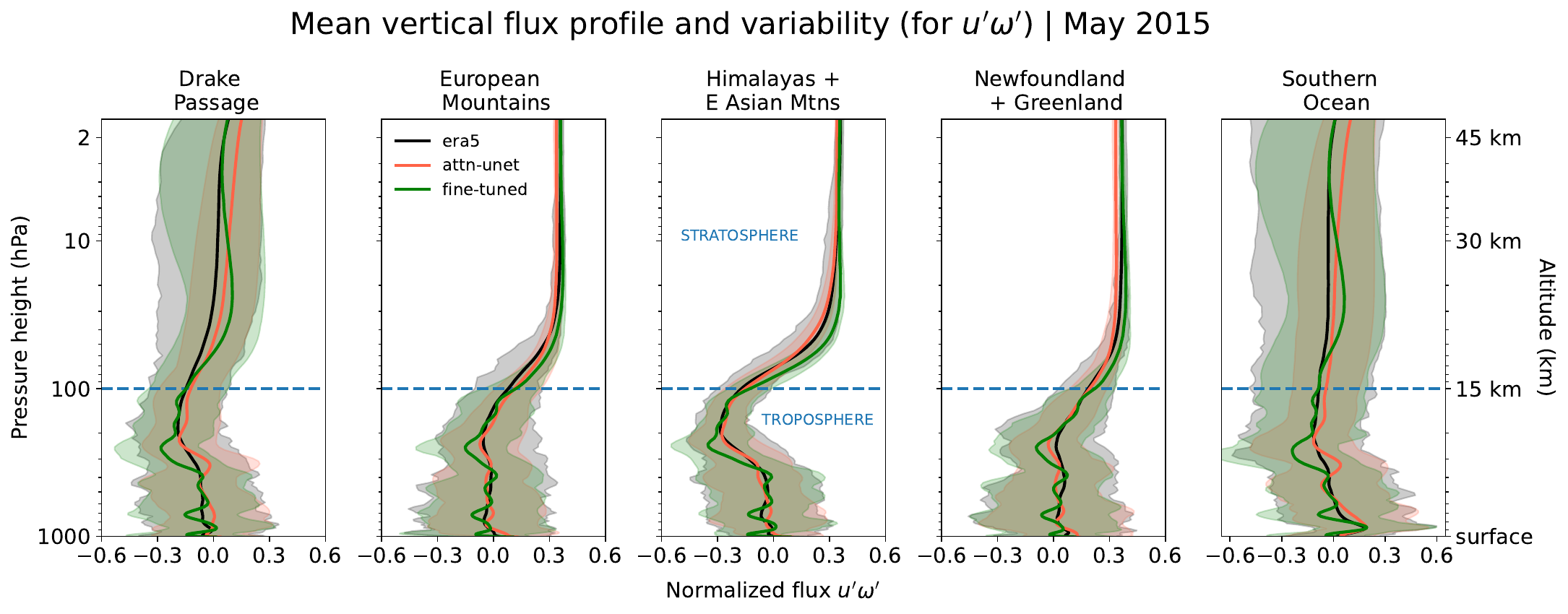}
\caption{May 2015 mean \change[R2]{true}{ERA5} and predicted vertical profiles of the normalized (unitless) zonal flux, $F_x=u'\omega'$ over five hotspots. The exact boundaries of the hotspots are shown in Fig \ref{fig:timeseries}. The (normalized) ERA5 flux is shown in black, the prediction from attn unet is shown in orange, and the prediction from the fine-tuned model is shown in green. The gray, orange, and green shadings show the range of flux variability in the respective models. \add[R3]{The regional extent for the hotspots is shown in Fig} \ref{fig:all_hotspots}. \add[R3]{The fluxes converge to near-zero in the stratosphere over all hotspots, due to successive wave filtering and dissipation, but appear to converge to non-zero values due to non-dimensionalization.}}
\label{fig:flux_profile1}
\end{figure}

While both models generate mean vertical profiles that are very similar to ERA5 over the five hotspots, the fine-tuned model generates both richer and more accurate variability in the stratosphere than the baseline (Fig \ref{fig:flux_profile1}). The difference in variability is substantial in the stratosphere. Both models generate weaker stratospheric variability than ERA5 in the summer stratosphere (\change[R1]{Scandinavia}{European mountains}, Himalayas, and Newfoundland) owing to weak GW activity in the region.\add[R1]{ The shading in }Fig \ref{fig:flux_profile1}\add[R1]{ shows the range of the true and predicted fluxes}. Yet, the wintertime \add[R3]{stratospheric} variability over the Drake Passage and Southern Ocean in the fine-tuned model is more consistent with ERA5. This is consistent with the lower variance noted for baseline predictions in the upper stratosphere. The observed and predicted mean vertical flux profiles and their variability over different hotspots for the meridional flux is shown in Fig \ref{fig:flux_profile2}.

\subsection{Global Averaged Flux Distribution}

\begin{figure}[!ht]
\centering
\includegraphics[width=\textwidth]{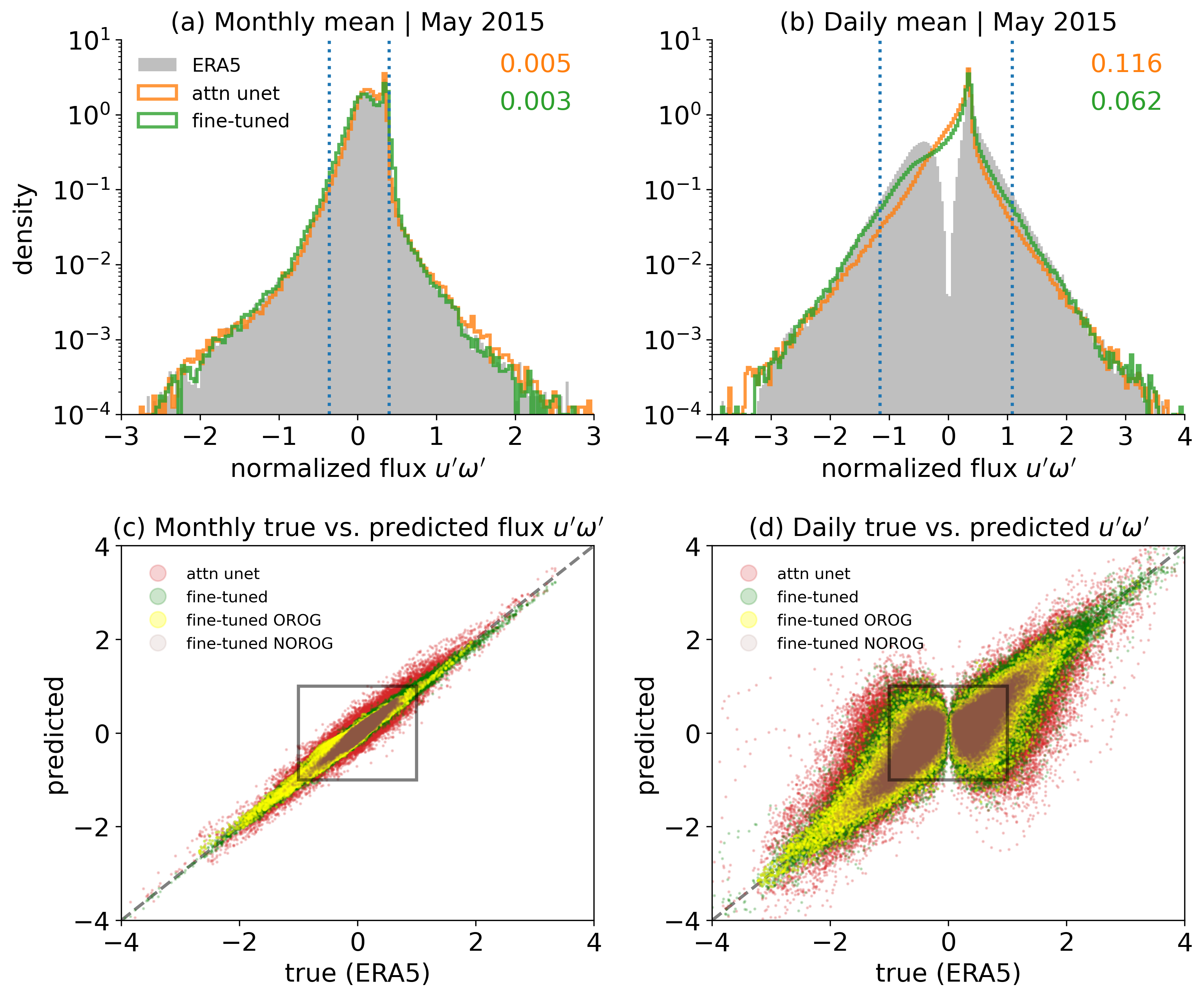} 
\caption{Histogram of the (a) May 2015 averaged and (b) daily averaged GW flux $u'\omega'$. Gray shading shows the underlying ERA5 distribution, orange is the attn unet prediction, and green is the fine-tuning prediction. Numbers indicate the Hellinger distance for the corresponding predictive model. The dotted lines show the 2.5$^{\textrm{th}}$ and 97.5$^{\textrm{th}}$ percentile, respectively (note the log-scaled y-axis). Scatter plot of the ERA5 vs. predicted flux \emph{for each point} for (c) May 2015 monthly average and (d) daily average. Red and green markers show the scatter for the baseline and fine-tuned model. The scatter for four orographic (Drake Passage, Himalayas, Newfoundland, and \change[R1]{Scandinavian}{European} Mountains) and nonorographic hotspots (tropical Pacific, North Atlantic, Southeast Asia, and Southern Ocean) for the fine-tuned model are shown in yellow and brown, respectively.  \add[R3]{The regional extent for the hotspots is shown in Fig} \ref{fig:all_hotspots}. \add[R3]{A gray box is added for reference over the [-1,1]x[-1,1] interval.}}
\label{fig:histogram}
\end{figure}

Global flux distributions provide insight into how well our models generate the possible range of flux responses globally, which are crucial to modeling extreme GW events. The observed and predicted global distribution of the GW momentum fluxes at different sampling frequencies (monthly vs. daily averaged) is shown in Fig \ref{fig:histogram} and \ref{fig:histogram_vw}. The histogram represents the distribution of the May 2015 monthly mean momentum flux globally, i.e., over all points in the troposphere and stratosphere. Both the baseline and the fine-tuned models simulate the monthly mean distribution with remarkable accuracy both in the bulk of the distribution and its tails (Fig \ref{fig:histogram}a). The baseline and the fine-tuned model have a Hellinger distance of 0.005 and 0.003 from the underlying training (ERA5) distribution, suggesting that despite clear differences in predictive skill, the two distributions are nearly identical to the underlying ERA5 distribution. The fine-tuned model emulates the distribution tails slightly better than the baseline.

\change[authors]{In a monthly climatology sense}{For monthly averages}, the fine-tuned model provides excellent prediction of the mean flux field, as is \remove[authors]{also }gauged by the scatter plot in \add[R3]{Fig} \ref{fig:histogram}c. The fine-tuned (green) model exhibits a\remove[R1]{ massively} reduced dispersion compared to \change[R1]{the baseline}{attn unet} (red). The fluxes over orographic hotspots (yellow) account for a \change[authors]{majority of the}{larger} scatter, \change[authors]{with}{and fluxes from} nonorographic hotspots (brown) \change[authors]{accounting for smaller values}{clustered around smaller values}.

Both models, however, struggle to accurately capture the daily-sampled histogram around small values. In the $\left[-0.5,0.5\right]$ interval, the models fail to accurately learn the small values and instead predict zero more frequently. This implies that the models learn the strong GW events more readily than the weak ones. \change[R3]{As a result of this inability, for daily sampling, the baseline and fine-tuned models have a degraded Hellinger distance of 0.116 and 0.062, respectively.}{Accordingly, the daily sampled flux distributions from both attn unet and the fine-tuned models produce higher Hellinger distances of 0.116 and 0.062, respectively.} \remove[R1]{This is reminiscent of prevailing problems with even state-of-the-art AI weather prediction models, which skillfully predict large-scale features but fail to reach the same level of accuracy in predicting the small scales.}

The attn unet model exhibits a visibly larger scatter between \change[R2]{the true}{ERA5} and predicted flux values compared to the fine-tuned model (Fig \ref{fig:histogram}d, red markers vs green markers). \change[R3]{As is the case for monthly sampling, the orographic hotspots account for most of the scatter dispersion while the nonorographic fluxes account for the smaller values around zero, which the model predicted more poorly. This partly explains why the correlation for both models is poorer in the tropics, where almost all GWs have nonorographic origins.}{The NOROG  data points (brown markers) exhibit less scatter in panels (c) and (d), implying a tighter clustering around the diagonal compared to OROG data points (yellow markers). This suggests that the small GW fluxes that the models struggle to capture (panel (b)) are likely dominated by NOROG sources, indicating that despite their tighter clustering, the NOROG points may still have significant systematic bias near zero.} \remove[R3]{This also explains why the predicted flux map in Fig} \ref{fig:intro}\remove[R3]{ looks smoother than the true flux map from ERA5.}

\section{Conclusion and Discussion}

Our analysis establishes that the atmospheric evolution learned by large transformer-based foundation models (developed for weather research) can be leveraged to \remove[R3]{simplify}, improve, and expedite the creation of subgrid-scale parameterizations for climate models. The \change[R1]{fine-tuned}{FM} parameterization \remove[R1]{ physically} \add[R1]{more accurately predicts lateral propagation effects} than traditional parameterizations and outperforms the \change[R1]{previous AI}{advanced attn unet} benchmark with fewer learnable parameters, even in upper atmospheric regions where the foundation model was not pre-trained. This provides a fresh avenue to develop AI-driven representation of small-scale processes that have the potential to replace traditional parameterizations, which for over four decades have been envisioned as single-column idealized sub-modules that often neglect key process physics. Coupling these parameterizations with existing climate models can promote and expedite the development of hybrid climate prediction models. 

Since Prithvi is trained on large amounts of data, its latent encoder-decoder space contains a rich abstract representation of atmospheric evolution. Since Prithvi's training data includes atmospheric variables like winds, humidity, radiation, and even leaf area index and soil moisture, the application of this approach transcends GWs and, with appropriate observations for fine-tuning, can be used to create parameterizations for other atmosphere-ocean-land processes. As an added benefit, this approach allows \change[R3]{blending}{using} data from multiple streams \add[R3]{for finetuning} --- \remove[R3]{synthetic }high-resolution model data, satellite trains, terrestrial remote sensing data, ground observations, etc.


In the context of atmospheric gravity waves, the \change[R1]{fine-tuned model outshines}{FM outperforms} the attn unet \change[R1]{baseline}{benchmark} \change[R3]{all three metrics: instantaneous prediction, regional distribution, and global distribution}{in representing GW effects in the upper stratosphere}. \change[R3]{In addition, i}{I}t learns the \add[R3]{effects of} three-dimensional propagation and dissipation of GWs in the atmosphere \add[R3]{similar to attn unet} \change[R3]{and generalizes}{while also generalizing} to regions unseen during \add[R1]{pre-}training. As a result, this model is capable of representing the missing GW effects in coarse-climate models. These effects are critical to getting a more realistic middle atmospheric circulation and seasonal wind transitions in climate models and in alleviating the cold-pole bias that undermines the accuracy of these models.

Recent studies, for instance, \citeA{Bretherton.etal2022} have discussed using deep learning models trained on high-resolution model simulations to ``bias-correct" coarse-climate models. While effective, these techniques are less interpretable, as it is challenging to learn the true source of prevailing biases and to distinguish structural model errors from model errors due to inaccurate physics. Given the versatility of FMs, our approach even allows the development of process-specific \add[R3]{emulators for representation of a suite of subgrid-scale processes, leading to bias-correction through a closer-to-observed data-driven representation}.

\add[R1]{To demonstrate the central idea, we have fine-tuned our models on limited years of ERA5 reanalysis data spanning a few variables. As noted above, the fluxes in ERA5 are not assimilated but model-generated, and owing to a 25 km resolution, ERA5 misses a broad range of mesoscale atmospheric GWs (with wavelengths shorter than 150-200 km). In addition, while the fluxes computed from ERA5 using Helmholtz Decomposition might almost exclusively reflect fluxes carried by GWs, in the troposphere, it could have some contributions from strong convective forcings in regions with strong precipitation} \cite{Alexander.etal2006, Wei.etal2022}. \add[R1]{As a consequence, there is much scope for improvements in the parameterization. These improvements can be achieved by fine-tuning on longer periods of high-resolution (high-fidelity) datasets, through better estimates of tropospheric GW fluxes, and by including more convection-related variables (for instance, humidity, diabatic heating, latent heat fluxes). This will be the focus of future work, where the fine-tuning will be accomplished using a kilometer-scale, high-resolution model output and an expanded feature set instead.}

\add[R1]{Admittedly, the nonlocal architecture adopted in this study does not align with the widely adopted column-based discretization used by climate models. Yet, work is underway to couple the nonlocal fine-tuned scheme to an atmospheric model (National Center for Atmospheric Research (NCAR)'s CAM7 model) and evaluate its online performance on atmospheric variability and generalizability on warming scenarios. To this end, in collaboration with the Institute of Computing for Climate Science at the University of Cambridge and NCAR, we have identified inbuilt pooling and discretization functions that make this coupling possible using \emph{ftorch}, without adding latency} \cite{Atkinson.etal2025, Chapman.Berner2025}. 

Nevertheless, foundational models open avenues to using multisource observations to facilitate not just AI-powered weather research but also climate research. Due to constraints on computing power, we are still \change[R1]{decades}{far} away from being able to run climate models (such as those participating in CMIP) multiple decades and centuries into the future at kilometer or sub-kilometer resolutions. This means climate prediction will continue to miss crucial sub-grid physics and will continue to rely on physical parameterizations of unresolved processes. 

We have demonstrated an appealing application of an existing foundation model in improving the sub-grid scale physics representation in a climate model. \add[R1]{These emulators are not just intended to be coupled to a climate model, but can also serve as standalone plugins to improve small-scale variability in AI-based weather models and other fine-tuned models.} In principle\remove[R1]{, despite finite-time blowups,}, FMs like Prithvi could be strategically applied to address a range of climate applications, including, but not limited to, heat wave prediction, land-use trend detection, and \change[R3]{climate event compositing}{cross-domain learning}. \add[R3]{This can be accomplished by fine-tuning the base FM on event-specific datasets, to enhance performance on rare events as, for instance, is discussed in} \citeA{Cui.etal2025}. \add[R3]{More precisely, during pre-training, the FM is pre-trained on global forecasts to optimize the global mean squared error. Parts of the FM, when retrained on event-specific data, using a specialized loss function, can be used to quickly re-train and develop specialized task-emulators.} Used in conjunction with other foundation models, such as Prithvi HLS \cite{Jakubik.etal2023}, this approach can be leveraged to create lightweight, fine-tuned models for key weather and climate applications, including the prediction of wildfires, hurricane storm surge, and regional heatwave impacts, potentially improving climate change preparedness. \remove[R1]{Efforts are currently underway to couple our fine-tuned scheme to a coarse-climate model and assess its online performance and feedback on the model dynamics.}

\clearpage
\appendix
\section{Additional Model Information and Meridional Flux Predictions}
\begin{figure}[!ht]
\centering
\includegraphics[scale=0.6]{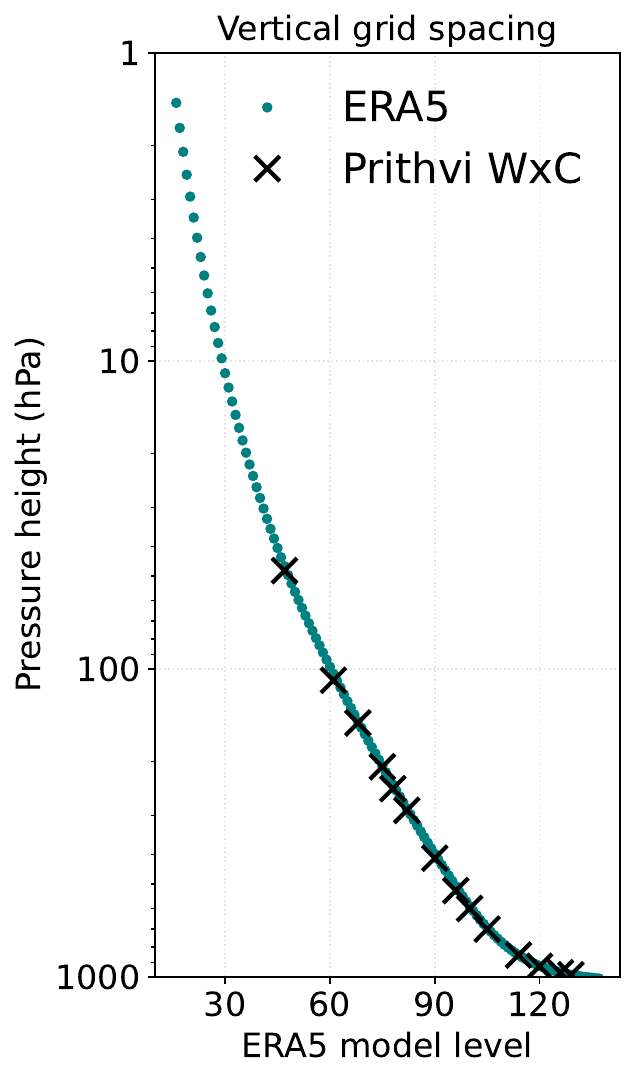}
\caption{Prithvi was pre-trained on very sparse data in the vertical. The ERA5 fine-tuned data was computed on 137 model levels, and the top 15 model levels (i.e., levels above 1 hPa $\sim$ 45 km) were discarded due to an artificial model sponge imposed at those levels. So, effectively 122 model levels between 1000 hPa (surface) to 1 hPa (45 km) height were used. In contrast, Prithvi is trained on MERRA-2 data interpolated to 14 vertical levels: [985, 970, 925, 850, 700, 600, 525, 412, 288, 245, 208, 150, 109, 48] hPa. No training data were provided between 50 hPa and 1 hPa during pre-training. This means that the frozen encoder-decoder has no prior knowledge about the dynamic evolution of gravity waves at these heights. Still, \change[R3]{as the analysis shows}{as shown in Figs }\ref{fig:hotspots_histogram} and \ref{fig:flux_profile1}, the fine-tuned model outperforms the baseline in this region.}
\label{fig:vert_spacing}
\end{figure}

\begin{figure}[!ht]
\centering
\includegraphics[width=1\textwidth]{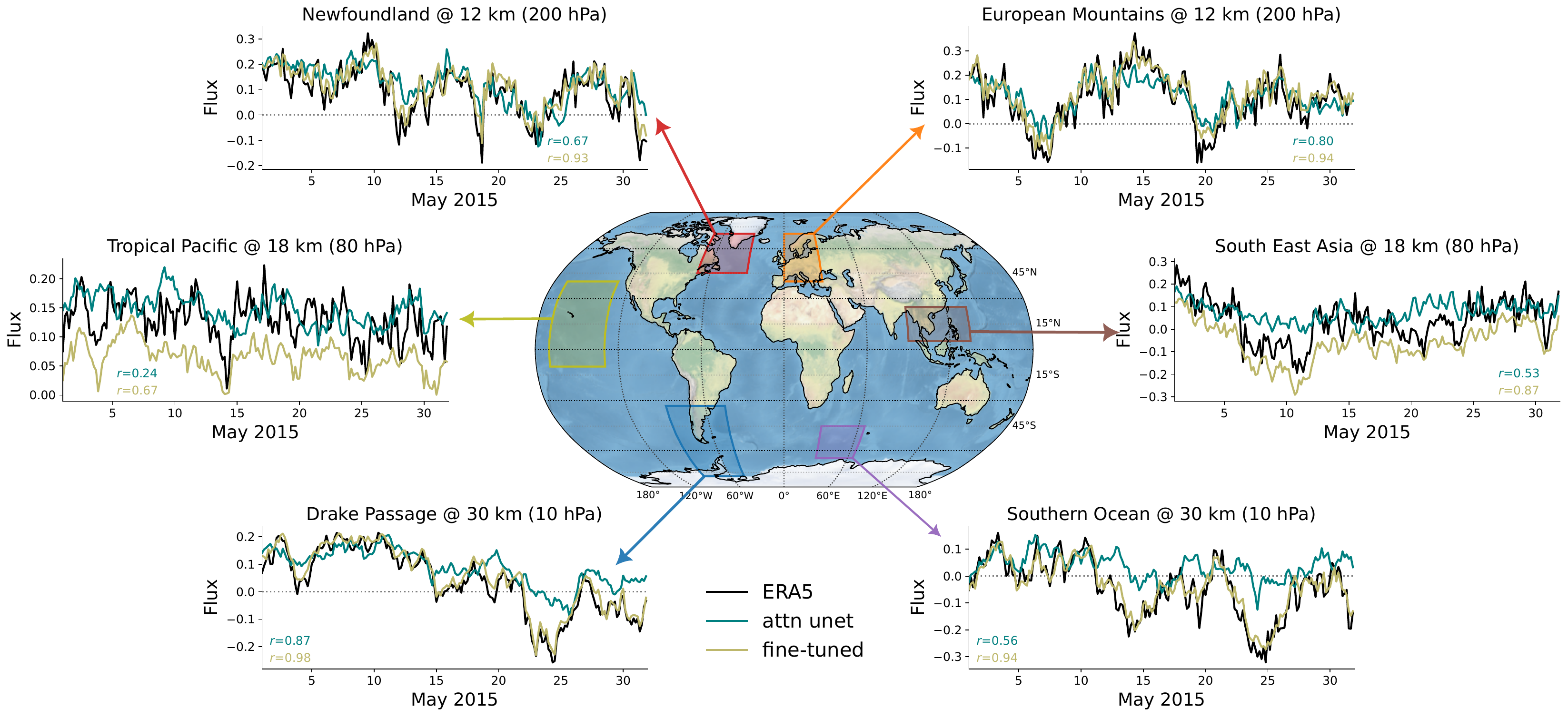} 
\caption{Same as Fig \ref{fig:timeseries} but for $v'\omega'$ - instantaneous fluxes for May 2015 from ERA5 (black), and predictions from the baseline (teal) and the fine-tuned model (light green) over six different hotspots.}
\label{fig:timeseries_vw}
\end{figure}

\begin{figure}
    \centering
    \includegraphics[width=0.5\linewidth]{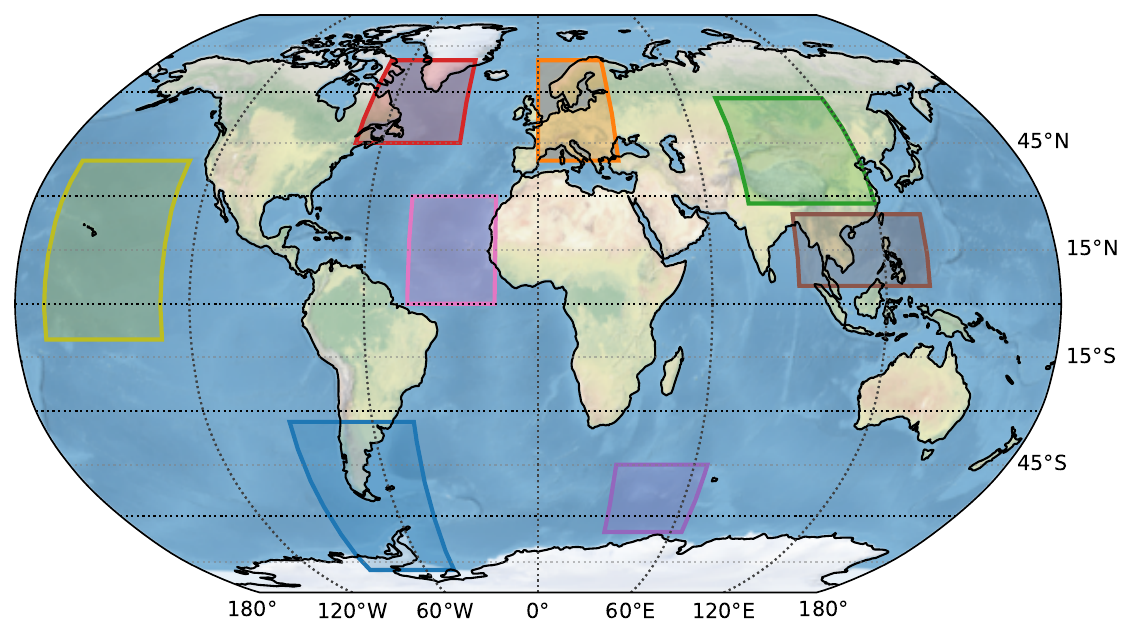}
    \caption{\add[R3]{All the GW hotspots considered for regional analysis in this work. Yelllow: Tropical Pacific (170$^{\circ}$W, 130$^{\circ}$W) $\times$ (10$^{\circ}$S, 40$^{\circ}$N), Red: Newfoundland Mountains + Southern Greeland (70$^{\circ}$W, 30$^{\circ}$W) $\times$ (45$^{\circ}$N, 70$^{\circ}$N), Orange: European Mountains (0$^{\circ}$, 30$^{\circ}$E) $\times$ (40$^{\circ}$N, 70$^{\circ}$N), Green: Himalayas and East Asian Mountains (75$^{\circ}$E, 120$^{\circ}$E) $\times$ (28$^{\circ}$N, 58$^{\circ}$N), Light Pink: Northern Atlantic (45$^{\circ}$W, 15$^{\circ}$W) $\times$ (0$^{\circ}$, 30$^{\circ}$N), Brown: Southeast Asia (90$^{\circ}$E, 135$^{\circ}$E) $\times$ (5$^{\circ}$N, 25$^{\circ}$N), Blue: Drake Passage (90$^{\circ}$W, 45$^{\circ}$W) $\times$ (78$^{\circ}$S, 33$^{\circ}$S), Dark Pink: Southern Ocean (30$^{\circ}$E, 65$^{\circ}$E) $\times$ (65$^{\circ}$S, 45$^{\circ}$S).}}
    \label{fig:all_hotspots}
\end{figure}


\begin{figure}[!ht]
\centering
\includegraphics[width=0.95\textwidth]{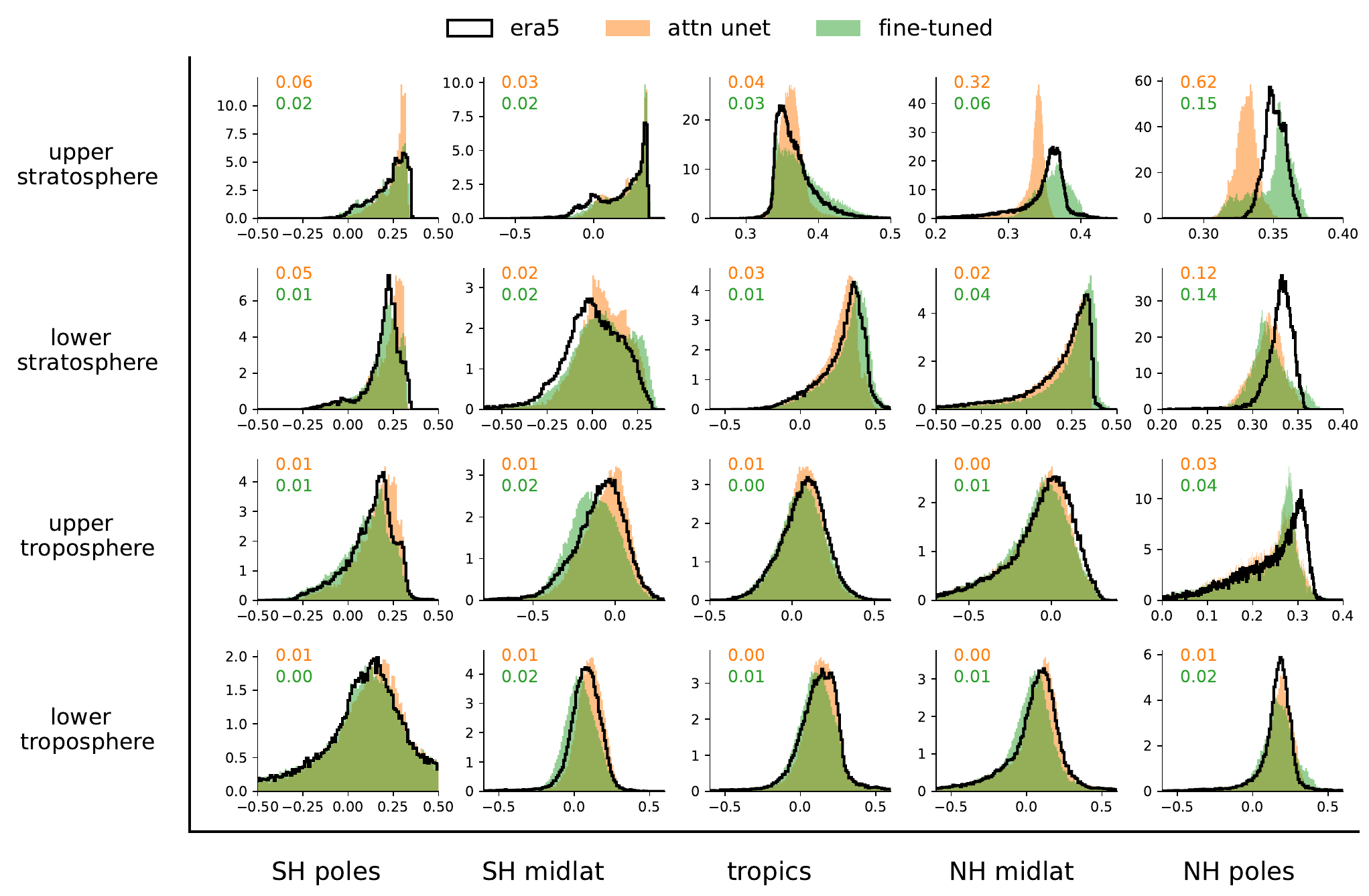} 
\caption{GW flux distributions similar to Fig \ref{fig:hotspots_histogram} but now divided according to latitude and height. The poles are defined as latitudes 60$^{\circ}$-90$^{\circ}$, the midlatitudes as 30$^{\circ}$-60$^{\circ}$, and the tropics as 15$^{\circ}$S-15$^{\circ}$N. The numbers indicate the respective Hellinger distances w.r.t. the \remove[R2]{true }distribution from ERA5 (black). For each latitude band, averaging is conducted over the whole latitude circle, i.e., over all longitudes.}
\label{fig:regional_histogram}
\end{figure}

The observed and predicted monthly mean distributions over different latitude bands are shown in Fig \ref{fig:regional_histogram}. The predicted and averaged fluxes agree quite well throughout the lower and upper troposphere. The Hellinger distances are consistently lower than 0.02 in most cases; an exception being the northern hemispheric poles in the upper troposphere. Within the troposphere, both models largely agree well on the captured distributions, as indicated by low Hellinger distances.

The Hellinger distances are higher throughout the stratosphere compared to the troposphere. Distances are less than 0.05 in the tropics and midlatitudes of the lower stratosphere, but get as high as 0.14 in the polar regions. The baseline performance, however, gets much worse in the upper stratosphere, with distances reaching up to 0.62 in the northern hemisphere upper stratosphere due to a notable shift bias in the predicted distribution. In contrast, the distances for the fine-tuned model remain below 0.15.

\begin{figure}[!ht]
\centering
\includegraphics[scale=0.35]{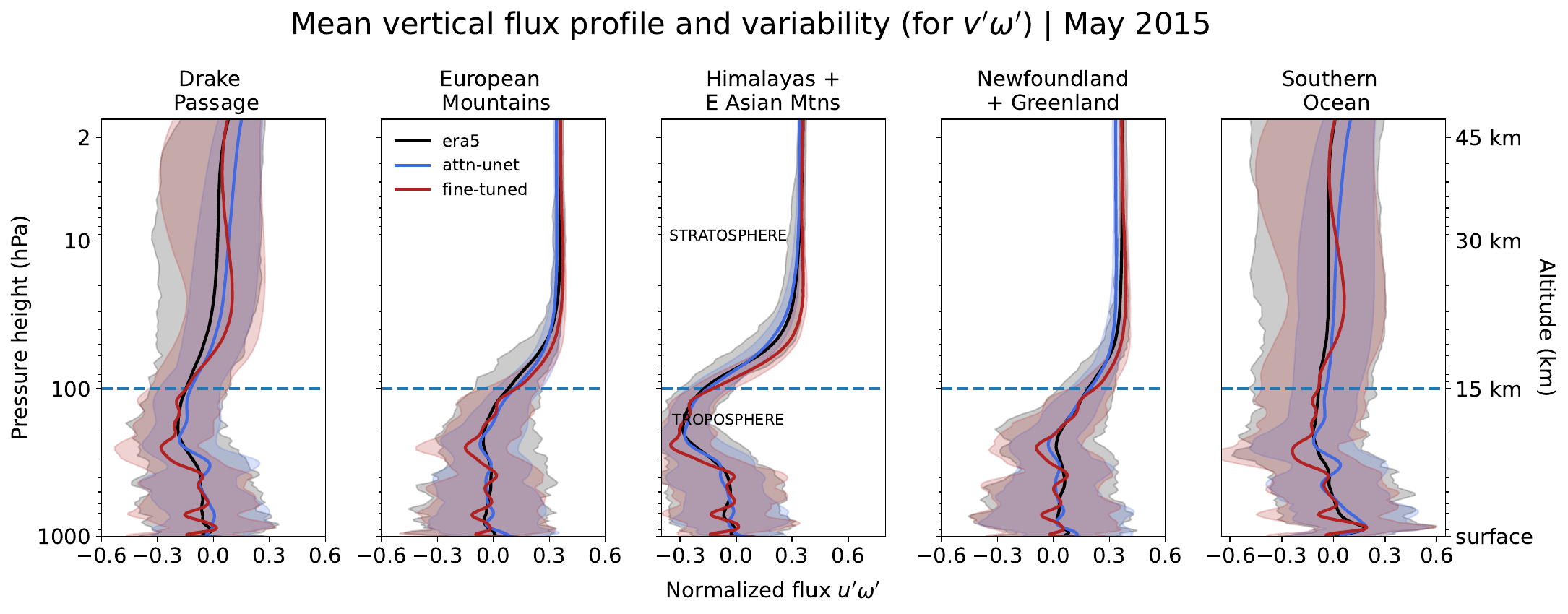}
\caption{Same as Fig \ref{fig:flux_profile1} but for the meridional flux $v'\omega'$.}
\label{fig:flux_profile2}
\end{figure}

\begin{figure}[!ht]
\centering
\includegraphics[width=\textwidth]{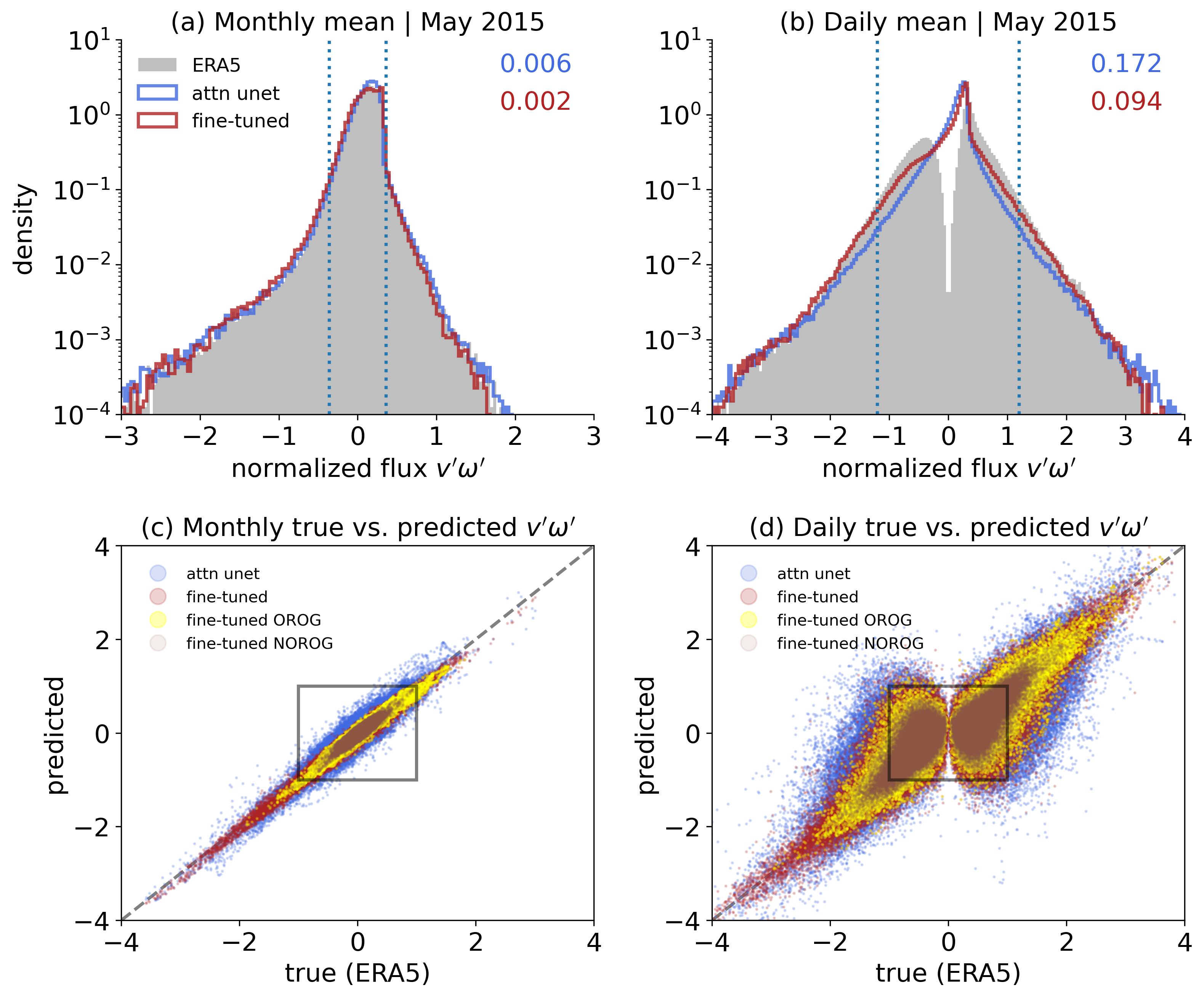} 
\caption{Same as Fig \ref{fig:histogram} but for $v'\omega'$. The distribution of the (a) May 2015 averaged and (b) daily averaged GW flux $v'\omega'$. Gray shading shows \remove[R2]{the true}{ERA5's} underlying distribution, orange is the baseline prediction, and green is the fine-tuning prediction. Numbers indicate the Hellinger distance for the corresponding predictive model. The dotted lines show the 2.5$^{\textrm{th}}$ and 97.5$^{\textrm{th}}$ percentile respectively (note the log-scaled y-axis). The bottom row shows the respective scatter of the \change[R2]{true}{ERA5} flux and the predicted meridional flux $v'\omega'$.}
\label{fig:histogram_vw}
\end{figure}
%



%
%
\clearpage
\section*{Open Research Section}

\begin{itemize}
\item \textbf{ERA5 data:} ECMWF’s ERA5 data \cite{Hersbach.etal2023} can be freely accessed from \url{https://cds.climate.copernicus.eu/datasets/reanalysis-era5-pressure-levels}. 
\item \textbf{ Prithvi WxC FM and fine-tuning:} The code for the Prithvi WxC model is available at \url{https://huggingface.co/ibm-nasa-geospatial/Prithvi-WxC-1.0-2300M}. The fine-tuning code for climate model parameterization for gravity wave flux is available at \citeA{RoyGupta2025}:\\ \url{https://doi.org/10.5281/zenodo.16666812}. Python scripts to compute gravity wave momentum fluxes from the publicly available ERA5 reanalysis are available at \citeA{Roy.etal2025}:\\ \url{https://doi.org/10.5281/zenodo.16666707}. 
\item \textbf{Python packages:} The default WindSpharm Python package is publicly available at \url{https://ajdawson.github.io/windspharm/}, and the PySpharm Python package is publicly available at: \url{https://pypi.org/project/pyspharm/}. The xESMF package used for conservative coarsegraining is publicly available at: \url{https://xesmf.readthedocs.io/en/stable/}.
\end{itemize}

\acknowledgments
Aditi Sheshadri and Aman Gupta are supported by Schmidt Sciences, LLC, as part of the Virtual Earth
System Research Institute (VESRI). Aditi Sheshadri also acknowledges support from the National Science Foundation through Grant OAC-2004492. The work was also supported by NASA’s Office of Chief Science Data Officer and Earth Science Division’s Earth Science Scientific Computing, Earth Science Data Systems Program, and the Earth Science Modeling and Analysis Program.

%
%

\bibliography{references}

%
%
%
%
%

\end{document}